# Positive and unlabelled machine learning reveals new fast radio burst repeater candidates

Arjun Sharma [1]★ and Vinesh Maguire Rajpaul [2]

[1]*The Shri Ram School, V-37, Moulsari Ave, Sector 24, Gurugram, Haryana 122002, India*
[2]*Isaac Newton Institute, University of Cambridge, 20 Clarkson Rd, Cambridge CB3 0EH, UK*

## ABSTRACT

Fast radio bursts (FRBs) are astronomical radio transients of unknown origin. A minority of FRBs have been observed to originate from repeating sources, and it is unknown which apparent one-off bursts are hidden repeaters. Recent studies increasingly suggest that there are intrinsic physical differences between repeating and non-repeating FRBs. Previous research has used machine learning classification techniques to identify apparent non-repeaters with repeater characteristics, whose sky positions would be ideal targets for future observation campaigns. However, these methods have not sufficiently accounted for the positive and unlabelled (PU) nature of the data, wherein true labels are only available for repeaters. Modified techniques that do not inadvertently learn properties of hidden repeaters as characteristic of non-repeaters are likely to identify additional repeater candidates with greater accuracy. We present in this paper the first known attempt at applying PU-specific machine learning techniques to study FRBs. We train an ensemble of five PU-specific classifiers on the available data and use them to identify 66 repeater candidates in burst data from the CHIME/FRB collaboration, 18 of which were not identified with the use of machine learning classifiers in past research. Our results additionally support repeaters and non-repeaters having intrinsically different physical properties, particularly spectral index, frequency width, and burst width. This work additionally opens new possibilities to study repeating and non-repeating FRBs using the framework of PU learning.

## 1 INTRODUCTION

Fast radio bursts (FRBs) are astronomical radio transients of unknown origin (Zhang 2023) characterized by their $\mu$s to ms-duration and high dispersion measures (Connor & Petroff 2018). A minority of FRB sources are known to repeat. Recent literature presents two possible theories to explain this: either all FRBs repeat, with repetition intervals varying significantly across sources, or perhaps some FRBs are intrinsically one-off events, originating from a distinct sub-population compared to repeaters (Lin et al. 2023).

No progenitor model is widely agreed upon as explaining the origin of all FRBs (Platts et al. 2019; Zhang 2023), however, magnetars continue to be considered the leading model (Gordon et al. 2023; Zhang 2023; Zhang et al. 2023). Cataclysmic models have additionally been proposed as sources for one-off FRBs, with some recent results (Moroianu et al. 2023) supporting the theory that some one-off FRBs may arise from binary neutron star mergers (Falcke & Rezzolla 2014). Another recently proposed model suggests that interactions between gravitational waves and pulsar magnetospheres may be responsible for both repeating and non-repeating FRBs (Kalita & Weltman 2023; Kushwaha, Malik & Shankaranarayanan 2024).

An increase in the number of known repeaters may significantly contribute to the development of a more complete understanding of the origin of FRBs. Since repeaters have typically been localized to host environments with greater success (Andersen et al. 2023), an increase in the understanding of FRB host galaxies caused by the identification and localization of new repeaters may contribute to a better understanding of the physical environments which produce FRB progenitors (Bhandari et al. 2022). Further analysis of the extent to which repeaters and non-repeaters differ in terms of their physical properties may additionally indicate whether they emerge from distinct processes.

The recent increase in the availability of data for observed bursts, particularly the release of the first CHIME/FRB catalogue (CHIME/FRB Collaboration 2021), has allowed extensive comparisons of repeaters and non-repeaters. Significant differences in the distributions of the burst widths and bandwidths of repeating bursts and apparently non-repeating bursts were identified by CHIME/FRB Collaboration (2021). Further work, such as Pleunis et al. (2021), Zhang et al. (2022), and Zhong et al. (2022), has since found differences in other observed properties of bursts, including extragalactic dispersion measure, peak frequency, and spectral running. This has lent increasing support to the hypothesis that bursts from repeating and non-repeating sources have intrinsically different physical characteristics, and has thus motivated the use of machine learning techniques (Pleunis et al. 2021; Chen et al. 2022; Luo, Zhu-Ge & Zhang 2023; Yang et al. 2023; Zhu-Ge, Luo & Zhang 2023) to provide additional insights into the differences between repeaters and apparent non-repeaters. Pleunis et al. (2021), Luo et al. (2023), and Zhu-Ge et al. (2023) identify *repeater candidates*: apparent non-repeaters which have a high probability of being repeating in nature. Accurate identification of candidates can increase the efficiency of future observing campaigns, which typically rely on follow-up observations of known burst sky positions (Connor & Petroff 2018).

★ E-mail: arjun.sharma07@outlook.com

FRB repeaters and apparent non-repeaters can be described as positive and unlabelled (PU) data. Some positive examples, i.e. known repeaters, are labelled, while apparent non-repeaters are all unlabelled. It is unknown which of the unlabelled examples are truly positive, i.e. hidden repeaters, and which are truly negative, i.e. non-repeaters.

However, past approaches to identifying repeater FRB candidates using supervised learning techniques, as in Luo et al. (2023), implicitly make the negativity assumption for PU data described by Bekker & Davis (2020), wherein all unlabelled examples are treated as belonging to the negative class. Classifiers trained under this assumption may learn the features of hidden repeaters as characteristic of non-repeaters, and therefore misclassify some hidden repeaters.

The development of modified classification approaches to overcome this challenge in PU data has been an active area of research for over two decades (Bekker & Davis 2020). In this paper, we introduce the first application of these PU-specific learning techniques to identify candidate FRB repeaters.

In Section 2, we characterize the FRBs and features we use in our analysis. In Section 3, we describe our approach for developing and evaluating PU classifiers. In Section 4, we analyse the performance of our classifiers and use them to identify repeater candidates, comparing our results with previous work. In Section 5, we discuss the implications of our work on the study of FRBs and future research possibilities.

## 2 DATA

### 2.1 Data sources

We use the first CHIME/FRB catalogue (CHIME/FRB Collaboration 2021), which contains bursts detected between 2018 July 24 and 2019 July 1 by the CHIME telescope in 400–800 MHZ range; as well as the bursts detected by CHIME between 2019 September 30 and 2021 May 1 (Andersen et al. 2023). We refer to these two data sets as the '2021 catalogue' and '2023 catalogue' respectively.

Each burst may contain one or more individual pulses within its dynamic spectrum, which are referred to as its sub-bursts (Brown et al. 2024).

The 2021 catalogue originally consisted of 600 sub-bursts, of which 506 are from apparently non-repeating sources, and 94 from 18 known repeating sources. We exclude the six sub-bursts from bursts that do not have flux measurements (FRB20190307A, FRB20190307B, FRB20190329B, FRB20190329C, FRB20190531A, and FRB20190531B), all of which are apparent non-repeaters, thus leaving 594 sub-bursts, of which 500 are from apparently non-repeating sources. In the context of PU learning, we treat known repeaters as positive examples and apparent non-repeaters as unlabelled examples.

The 2023 catalogue consists of 151 sub-bursts from 127 bursts, all of which have flux measurements. Here, for each cluster of bursts that are identified as likely to have originated from the same repeating source, Andersen et al. (2023) calculate the contamination rate $R_{cc}$, the expected number of false positives (FPs) in that cluster. 98 bursts, which comprise 119 sub-bursts, originating from 25 repeating sources with $R_{cc} < 0.5$ are treated as repeaters. Another 14 clusters with $0.5 \leq R_{cc} < 5$ are described as the 'silver sample' by Andersen et al. (2023). This contains 29 bursts composed of 32 sub-bursts. We treat these as unlabelled examples in our experiments. We merge the two data sets for our experiments. A description of the composition of the original and merged data sets is presented in Table 1.

**Table 1.** Distribution of FRB data in the 2021 catalogue, 2023 catalogue, and the merged data set used in experiments. The discrepancy in the row for unlabelled data arises from seven bursts and seven sub-bursts from the 2021 catalogue being considered apparent non-repeaters at the time but later being identified as repeaters in the 2023 catalogue. These values were omitted from the unlabelled category and only treated as repeaters in the combined set.

|  |  | 2021 | 2023 | Overlap | Merged |
|---|---|---|---|---|---|
| All | Bursts | 530 | 127 | 14 | 643 |
|  | Sub-bursts | 594 | 151 | 15 | 730 |
| Positive | Bursts | 62 | 98 | 0 | 160 |
|  | Sub-bursts | 94 | 119 | 0 | 213 |
|  | Repeating sources | 18 | 25 | 0 | 43 |
| Unlabelled | Bursts | 468 | 29 | 7 | 483 |
|  | Sub-bursts | 500 | 32 | 8 | 517 |

### 2.2 Features

The input features used for our classifiers consist of primary characteristics directly provided in the CHIME catalogue and secondary features derived from known properties of the bursts. In Table 2, we list the nine physical characteristics observed by CHIME which we use as input features.

It was reported by CHIME/FRB Collaboration (2021) that the source density of the CHIME telescope is higher near the North Celestial pole, resulting in confusion which makes repeater identification more difficult than at lower declinations. Therefore, we use right ascension as an input feature but not declination.

We follow the method described by Luo et al. (2023) to estimate the redshift $z$ and luminosity distance $D_L$ of each burst. Using these, we calculate the same four secondary features as Luo et al. (2023) for each burst, which we briefly describe in Table 3. Considering differences in the values of features across sub-bursts, we follow previous work (Pleunis et al. 2021; Zhang et al. 2022; Zhong et al. 2022) and treat each sub-burst as a separate training example in our experiments, allowing for the possibility that features specific to the sub-bursts of a burst may encode information about the repeating nature of a source.

## 3 METHODS

The methods described in this section are the same as in a previous paper (Sharma 2023) applying PU learning to the first catalogue.

### 3.1 Evaluating performance on PU data

We consider known repeaters to be a set of known positive examples and apparent non-repeaters to be a set of unlabelled examples which consist of subsets of hidden positive and hidden negative examples. Since there are available positive examples, the number of true positive (TP) and false negative predictions (FN) of a classifier on this data can be calculated. We cannot identify FP or true negative (TN) predictions, due to a lack of known negative examples. We therefore use two criteria to evaluate classifier performance on PU data.

---

[1] Declination was later tested with no meaningful difference observed on the performance of the classifier as presented in Section 4

**Table 2.** Primary input features. For each feature, the 'log' column notes whether its base-10 logarithm values are taken to make the distribution of values less skewed. We also note whether all sub-bursts (SBs) of a particular burst share the same value for each feature, as specified by CHIME/FRB Collaboration (2021).

| Feature | Description | Unit | Log | SBs |
|---|---|---|---|---|
| Right ascension | Sky position as per the J2000.0 equinox | ° | ✗ | ✓ |
| Signal-to-noise ratio | SNR computed by the CHIME `fitburst` fitting algorithm based on the strength of the entire FRB event | ... | ✗ | ✓ |
| Extragalactic DM | DM after subtracting the expected maximum Galactic contribution towards the source as per NE2001 | pc cm$^{-3}$ | ✓ | ✓ |
| Boxcar width | Estimate of the total duration of the entire burst | ms | ✓ | ✓ |
| Flux | Peak flux in the dynamic spectrum of the burst across all its sub-bursts | Jy | ✓ | ✓ |
| Fluence | Integral of the flux of the burst | Jy ms | ✓ | ✓ |
| Spectral index | Measure of the dependence of flux on frequency for each sub-burst | ... | ✗ | ✗ |
| Spectral running | Dependence of the spectral index on frequency | ... | ✗ | ✗ |
| Peak frequency | Frequency of each sub-burst at its highest flux density | MHz | ✓ | ✗ |

**Table 3.** Secondary input features. We take the base-10 logarithm values of all features. None have the same value for all sub-bursts of a burst, as specified by CHIME/FRB Collaboration (2021).

| Feature | Description | Unit | Log | SBs |
|---|---|---|---|---|
| Rest-frame width | Observed width of each sub-burst, as determined by the fitting algorithm `fitburst`, corrected for time dilation | s | ✓ | ✗ |
| Rest-frame frequency width | Difference between the highest and lowest observed frequencies of the burst, corrected for time dilation | MHz | ✓ | ✗ |
| Brightness temperature | Thermodynamic temperature of a blackbody that would emit radiation of an equivalent intensity | K | ✓ | ✗ |
| Burst energy | Upper limit of the isotropic energy of the burst within the observed bandwidth | erg | ✓ | ✗ |

**Recall**: the recall $r$ of a classifier is calculated as

$$r = \frac{TP}{TP + FN}. \tag{1}$$

**$L^2$ score (Lee-Liu score)**: Let the precision $p$ of a classifier be defined as

$$p = \frac{TP}{TP + FP}. \tag{2}$$

Then, Lee & Liu (2003) found that, for a binary classifier with true precision $p$ and recall $r$, whose output on a given example with true label $y$ is given by $\hat{y}$, the following expression is true:

$$\frac{pr}{P(y=1)} = \frac{r^2}{P(\hat{y}=1)}. \tag{3}$$

$P(y = 1)$ represents the probability of an example in the set being positive, i.e. a repeater; and $P(\hat{y} = 1)$ represents the probability of the output of the classifier being positive, i.e. the probability of the classifier flagging an example as a repeater. Thus, both $r$ and $P(\hat{y} = 1)$ can be calculated with the predictions of a classifier on PU data. Therefore, the right-hand side term in equation (3) can be directly calculated. We refer to this value as the $L^2$ score. It is large when both $p$ and $r$ are large, and small when one of $p$ or $r$ is small. Therefore, it can be interpreted similar to the conventional $F$-score metric. However, unlike $F$-score, $L^2$ score has an infinite range. Therefore, it can be used to compare the relative performance of two classifiers on the same set of data, where a higher $L^2$-score corresponds to more accurate predictions, but it cannot be interpreted in isolation.

### 3.2 Standard supervised classifiers

Three of the five PU classifiers used (described in Section 3.3) require, as a basis for their predictions, a supervised classifier that predicts the probability of an example being labelled. Therefore, to identify the optimal base classification techniques for the PU classifiers, we evaluate the performance of eight conventional (non-PU) supervised classifiers when trained using the available labels in the data.

#### 3.2.1 List of supervised classifiers tested

(i) **Decision tree**: uses a series of simple decision rules, based on input values, to predict a label (Breiman 2017).

(ii) **Random forest**: an ensemble of decision trees are fitted on random sub-samples of the training set using a bagging method. The final prediction is the average output of the ensemble (Breiman 2001).

(iii) **Support vector machine (SVM)**: projects input features into a higher dimensional plane using a kernel function, and attempts to find a hyperplane separating them (Boser, Guyon & Vapnik 1992).

(iv) **AdaBoost**: boosting ensemble method which trains an initial classification ensemble on the data, then gives higher weight to misclassified examples in the next training iterations (Freund & Schapire 1996).

(v) **Extreme gradient boosting (XGBoost)**: boosted ensemble technique that sequentially constructs decision trees in a depth-wise manner, optimizing a loss function and using regularization to prevent overfitting (Chen & Guestrin 2016).

(vi) **LightGBM**: boosted ensemble method that grows its decision trees leaf-wise, where the leaf with the largest loss is grown (Ke et al. 2017).

(vii) **Logistic regression (LR)**: predicts a probability of correspondence with one of two possible classes by applying the logistic

**Table 4.** Optimal hyperparameters for standard supervised classifiers found across 100 trials. All parameter values were sampled over a linear domain, unless specified.

| Model | Parameter | Values tested | Optimal value |
|---|---|---|---|
| SVM | `C` | 0.010 to 100 (`log`) | 8.472 |
| | `degree` | 1 to 8 | 5 |
| LDA | `solver` | `svd`, `lsqr`, `eigen` | `lsqr` |
| | `store_covariance` | `True`, `False` | `True` |
| | `tol` | $10^{-5}$ to $10^{-3}$ (`log`) | $2.05 \times 10^{-5}$ |
| LR | `tol` | $10^{-5}$ to $10^{-3}$ (`log`) | $5.62 \times 10^{-5}$ |
| | `C` | 0.010 to 100 (`log`) | 63.500 |
| | `solver` | `liblinear`, `newton-cholesky` | `liblinear` |
| | `max_iter` | 100 to 500 | 162 |
| Random forest | `n_estimators` | 50 to 500 | 245 |
| | `min_samples_split` | 2 to 32 | 21 |
| | `min_samples_leaf` | 1 to 32 | 10 |
| | `criterion` | `gini`, `entropy` | `entropy` |
| AdaBoost | `n_estimators` | 50 to 500 | 379 |
| | `learning_rate` | $10^{-3}$ to 1 | 0.280 |
| | `algorithm` | `SAMME`, `SAMME.R` | `SAMME.R` |
| XGBoost | `n_estimators` | 50 to 500 | 472 |
| | `eta` | $10^{-3}$ to 10 (`log`) | 4.46 |
| | `gamma` | $10^{-3}$ to 10 (`log`) | $1.58 \times 10^{-2}$ |
| | `min_child_weight` | $10^{-3}$ to 10 (`log`) | $1.32 \times 10^{-3}$ |
| | `max_delta_step` | $10^{-3}$ to 10 (`log`) | $2.15 \times 10^{-2}$ |
| | `max_leaves` | 2 to 256 | 165 |
| | `max_bin` | 2 to 256 | 81 |
| | `subsample` | 0.1 to 1 | 0.809 |
| | `colsample_bytree` | 0.1 to 1 | 0.426 |
| LightGBM | `n_estimators` | 50 to 500 | 314 |
| | `learning_rate` | $10^{-3}$ to 1 (`log`) | 0.763 |
| | `subsample` | 0.1 to 1 | 0.552 |
| | `colsample_bytree` | 0.1 to 1 | 0.875 |
| Decision Tree | `min_samples_split` | 2 to 32 | 20 |
| | `min_samples_leaf` | 1 to 32 | 2 |
| | `criterion` | `gini`, `entropy` | `gini` |

function to the weighted sum of the input features, mapping the value to be between 0 and 1 (Nelder & Wedderburn 1972).

(viii) **Linear discriminant analysis (LDA)**: projects data into a lower dimensional space which maximizes variance between the two classes. Then, it minimizes the variance within each class (Tharwat et al. 2017).

All classifiers are implemented using the SCIKIT-LEARN PYTHON framework (Pedregosa et al. 2011), except for XGBoost (Chen & Guestrin 2016) and LightGBM (Ke et al. 2017).

### 3.2.2 Optimization of supervised classifiers

The procedure followed to optimize each of the supervised classifiers is as follows.

(i) Data from the entire catalogue are randomly split into training and validation sets in an 8:2 ratio. In each set, we retain the same ratio of repeaters to non-repeaters as in the original data. For each burst, we ensure that all its sub-bursts are either in the validation set or the training set.[2]

(ii) We fit a standard distribution scaler using the SCIKIT-LEARN package (Pedregosa et al. 2011) on the training set, which, for each feature, removes the mean and scale values to unit variance. The fitted scaler is then used to transform the validation set.

(iii) We apply the synthetic minority oversampling technique (SMOTE) with the IMBALANCED-LEARN library (Lemaître, Nogueira & Aridas 2017) to balance the ratio of repeaters to apparent non-repeaters in the training set by synthesizing additional training examples of repeaters, based on the set of known repeaters. This is necessary for conventional supervised classifiers to prevent them from being biased by the ratio of known repeaters in the catalogue.

(iv) Each classifier is tuned with the OPTUNA hyperparamter optimization framework (Akiba et al. 2019). We identify some hyperparameters of each classifier, as listed in Table 4, as tunable, and define a space of possible values for each. Then, the Tree-structured Parzen Estimator (TPE) technique (Bergstra et al. 2011) is used to search for the hyperparameter configuration that maximizes the $L^2$ score of each classifier on the validation data after being fitted on the training set. Hyperparameter configurations are evaluated for each which share six input features – in the validation phase. This addresses what is likely an oversight in past work, such as Luo et al. (2023), that might have caused model performance to be overestimated.

---

[2]This prevents classifiers from being exposed to some sub-bursts of a burst in the training phase, and then predicting on the other sub-bursts of that burst –

classifier over 100 trials on the same fixed training and validation split. The combination which yields the best $L^2$ score is saved.

For each classifier, steps 1–4 are run, with the optimal configuration of hyperparameters found at the end of step 4 being saved. Then, to evaluate the average performance of the optimal version of each classifier, we repeat steps 1–3 1000 times for each classifier using its optimal hyperparameters. We save the $r$ and $L^2$ scores for each classifier when predicting on the validation set.

### 3.3 PU classifiers

Let an FRB example be denoted by $x$, and $s$ be the available label for that burst, which has a value of 1 for a labelled burst, i.e. a known repeater and 0 for an unlabelled burst, i.e. an apparent non-repeater. The true nature of the burst as either a repeater or non-repeater is given by $y$, which is unknown and is 1 for a true repeater and 0 for a true non-repeater. The class prior $\alpha = P(y = 1)$ represents the fraction of all FRBs that are repeaters. The value $c = P(s = 1|y = 1)$ represents the label frequency, i.e. the fraction of true repeaters in the data that have been correctly labelled as repeaters.

Bekker & Davis (2020) showed that, if it is assumed that the population of positive examples has been selected completely at random (SCAR) from the population of all positive examples in the data set, then $c$ can be defined as

$$c = \frac{P(s = 1)}{\alpha}. \tag{4}$$

This assumption allows us to apply five previously developed classifiers which utilize these conventions of PU learning to make predictions on PU data.[3]

#### 3.3.1 Classic Elkanoto (CE)

Elkan & Noto (2008) established a process by which the output of a standard supervised classifier could be modified for the context of PU classification.

First, a standard classifier is trained on the available label, $s$, of each training example $x$. If its probabilistic output on a new example $x$ is given by $g(x)$, then $g(x) \approx P(s = 1|x)$, i.e. the classifier will learn to predict the probability of the burst being labelled.

Then, $g(x)$ can be used to estimate $c$, the constant probability that a positive example is labelled in this data set, mathematically defined as $P(s = 1|y = 1)$. Assuming that $g(x)$ is exactly equal to $P(s = 1|x)$, an approximation of $c$ can be calculated using the subset of positive examples $P$ in a 'hold-out' set of $n$ examples randomly drawn in a fixed ratio from the training set. $c$ can be calculated as the average value of $g(x)$, and thus $P(s = 1)$, over all of these positive examples. Formally, this is stated as

$$c = \frac{1}{n} \sum_{x \in P} g(x). \tag{5}$$

Finally, if the SCAR assumption is true, Elkan & Noto (2008) showed that the probability of an example being labelled can be converted into the probability of the example being positive with the following equation:

$$P(y = 1|x) = \frac{P(s = 1|x)}{c}. \tag{6}$$

We refer to this classification technique as Classic Elkanoto (CE). CE requires the standard classifier to be well-calibrated, such that its predicted probabilities $g(x)$ can be interpreted as its confidence level in the prediction. Of the eight base classifiers in Section 3.2, only LR and LDA are considered to be well-calibrated algorithms by this definition. Therefore, wherever we use an uncalibrated base classifier, we manually calibrate it using Platt scaling over 10 cross-validation folds using Pedregosa et al. (2011).

#### 3.3.2 Weighted Elkanoto (WE)

It was shown by Elkan & Noto (2008) that, with the SCAR assumption,

$$P(y = 1|x, s = 0) = \frac{1 - c}{c} \frac{P(s = 1|x)}{1 - P(s = 1|x)}. \tag{7}$$

Therefore, the paper proposes giving labelled examples unit weight and duplicating unlabelled examples, with one copy being treated as a positive example with weight $P(y = 1|x, s = 0)$, and the other as a negative example with a complementary weight $1 - P(y = 1|x, s = 0)$.

Using equation (7), we can use a standard classifier $g(x)$ and a value of $c$ derived from the same method as CE to calculate the weights. As in CE, we apply Platt scaling where necessary.

#### 3.3.3 Bagged classifiers (BC)

Mordelet & Vert (2014) proposed a method in which multiple biased SVM classifiers are trained on all positive examples and randomly selected subsets of the unlabelled examples, with the unlabelled examples being treated as negative examples. The final output is the average of the predictions of all the classifiers. Mordelet & Vert (2014) proposed that this approach is likely to improve the performance of the classifier since, by coincidence, some subsets of the unlabelled examples are likely to be contaminated with fewer positive examples, so classifiers trained on those subsets are likely to be able to more accurately distinguish between positive and negative examples. The spatial variability in the trained classifiers can be exploited by the bagging procedure. We generalize this method such that it can be used for any of our base classifiers.

#### 3.3.4 Modified logistic regression (MLR)

We define the standard logistic regression (SLR) equation as

$$g_{LR(x)} = \frac{1}{1 + e^{-\hat{w} \cdot \hat{x}}}. \tag{8}$$

This has an upper bound value of 1. For $P(y = 1|x)$ to be a well-defined probability $0 < P(y = 1|x) < 1$ in equation (6), Jaskie, Elkan & Spanias (2019) establishes that $g(x) = P(s = 1|x) \leq c$. For this to be true, $P(s = 1|x)$ must lie in the range $[0, c]$, rather than in $[0, 1]$, as it does with SLR. The authors therefore introduce a modified logistic regression function that forces an upper bound estimate $\hat{c}$ of $c$ on the logistic regression equation, where

$$g_{MLR(x)} = \frac{1}{1 + b^2 + e^{-\hat{w} \cdot \hat{x}}}. \tag{9}$$

---

[3]In practice this assumption is reasonable though imperfect. For example, known CHIME selection biases mean that it may be more difficult to observe follow-up bursts from sources with high scattering frequencies and/or high dispersion measures (CHIME/FRB Collaboration 2021). Brighter (higher fluence) repeat bursts may also be easier to detect.

The new upper bound $\hat{c}$ is then equal to $\frac{1}{1+b^2}$ (Jaskie et al. 2019). They modify the learning process for SLR to find, in addition to the conventional weight vectors $\bar{w}$, the value of $b$.

### 3.3.5 PU Extra Trees (PUET)

Wilton et al. (2022) proposed a modified random forest approach for PU data. Random samples of PU data are extracted and decision trees are trained for each sample, with the optimal split of each tree chosen with a recursive greedy risk minimization approach based on PU-specific empirical risk estimation functions, such as the non-negative PU risk estimator (Kiryo et al. 2017) and unbiased PU risk estimator (Plessis, Niu & Sugiyama 2015).

PUET requires an estimation of the class prior $\alpha$ as an input. To calculate it, we estimate $c$ using a fitted instance of a CE classifier on the same training data, and use equation (4) to derive the corresponding value of $\alpha$.

### 3.3.6 Optimization of PU classifiers

The procedure used to train and optimize them is as below.

(i) The three classifiers from the previous stage that were found to have the highest mean recall score across all trials are selected as 'base classifier options', using their optimal hyperparameter configurations.

(ii) We follow the same procedure as in the previous phase to perform a training-validation split of the data in an 8:2 ratio and to apply standard scaling to the features. We do not apply any oversampling.

(iii) We fit each PU-specific classifier on the training set and use TPE to search for the hyperparameter configuration that maximizes the $L^2$ score of the classifier on the fixed validation set. For CE, WE, and BC, we treat the choice of base classifier as a tunable parameter using the base classifier options. The configuration that yields the best $L^2$ score across 150 trials, on the same fixed training and validation split, is saved.

To identify potential repeater FRB candidates in the catalogue, we then fit each PU classifier with its optimal configuration (as found in step 3) on a training set of 80 per cent of the data, and extract predictions from each classifier on the entire catalogue. We run this process 1000 times, using a different random training set in each iteration. We keep track of the number of iterations in which each unlabelled burst was flagged as a repeater by three or more classifiers. We consider bursts flagged more than 100 times to be candidate repeaters.

## 4 RESULTS

### 4.1 Classifier optimization

#### 4.1.1 Supervised classifiers

In Table 4, we list the hyperparameters tested for each supervised classifier, and the optimal values found. Other parameters are left at their default values. A complete list of all parameters used for each of the supervised classifiers is provided in Appendix A.

In Table 5, we list the mean values and standard deviations (SD) of the recall and $L^2$ score of the tuned versions of each of the supervised classifiers across 1000 trials, as per the procedure described in Section 3.2. SVM, LDA, and LR have the highest recall values;

**Table 5.** Average performance of optimized supervised classifiers across 1000 train-test splits, sorted by the mean value of $r$.

| Model | Mean $r$ | $r$ SD | Mean $L^2$ | $L^2$ SD |
|---|---|---|---|---|
| **SVC** | 0.845 | $5.69 \times 10^{-2}$ | 2.07 | 0.218 |
| **LDA** | 0.843 | $5.76 \times 10^{-2}$ | 2.09 | 0.219 |
| **LR** | 0.833 | $6.08 \times 10^{-2}$ | 2.06 | 0.225 |
| Random forest | 0.823 | $6.06 \times 10^{-2}$ | 2.08 | 0.221 |
| AdaBoost | 0.813 | $6.57 \times 10^{-2}$ | 2.09 | 0.240 |
| XGBoost | 0.809 | $6.35 \times 10^{-2}$ | 2.21 | 0.251 |
| LightGBM | 0.799 | $6.57 \times 10^{-2}$ | 2.15 | 0.248 |
| Decision tree | 0.696 | $8.58 \times 10^{-2}$ | 1.60 | 0.280 |

**Table 6.** Optimal hyperparameters for PU classifiers found across 150 trials.

| Model | Parameter | Values tested | Optimal value |
|---|---|---|---|
| CE | Base classifier | LDA, SVM, LR | LDA |
| | Holdout ratio | 0.1 to 0.8 | 0.673 |
| WE | Base classifier | LDA, SVM, LR | LR |
| | Holdout ratio | 0.1 to 0.8 | 0.387 |
| BC | Base classifier | LDA, SVM, LR | SVM |
| | Estimators | 25 to 200 | 156 |
| | Maximum samples | 0.100 to 1 | 0.610 |
| | Maximum features | 0.100 to 1 | 0.977 |
| MLR | Learning rate | $10^{-4}$ to $10^{-1}$ | $1.57 \times 10^{-2}$ |
| PUET | Estimators | 25 to 200 | 95 |
| | Risk estimator function | uPU, nnPU | uPU |
| | Loss function | Quadratic, logistic | Quadratic |
| | Minimum samples for leaf node | 1 to 10 | 3 |
| | Features used out of $n$ | $\sqrt{n}$ and $n$ | $\sqrt{n}$ |
| | Split points per candidate feature | 1 to 10 | 5 |

**Table 7.** Average performance of optimized PU classifiers across 1000 train-test splits.

| Model | Mean $r$ | $r$ SD | Mean $L^2$ | $L^2$ SD |
|---|---|---|---|---|
| CE | 0.842 | $1.87 \times 10^{-2}$ | 2.12 | $4.20 \times 10^{-2}$ |
| WE | 0.903 | $3.71 \times 10^{-2}$ | 2.02 | $8.49 \times 10^{-2}$ |
| BC | 0.921 | $6.55 \times 10^{-3}$ | 2.01 | $2.55 \times 10^{-2}$ |
| MLR | 0.735 | $3.6573 \times 10^{-2}$ | 2.03 | $7.20 \times 10^{-2}$ |
| PUET | 0.987 | $7.163 \times 10^{-3}$ | 2.03 | $8.71 \times 10^{-2}$ |

therefore, as per the criteria defined in Section 3.3.6, we use them as the base classifier options for CE, WE, and BC in the next stage.

#### 4.1.2 PU classifiers

In Table 6, we list the hyperparameters tested for each PU classifier, and the optimal values found. Other parameters are left at their default values. A complete list of all parameters used for each of the PU classifiers is provided in Appendix B.

In Table 7, we list the mean values and standard deviations of the recall and $L^2$ score of the tuned versions of each of the PU classifiers across 1000 trials, as per the procedure described in Section 3.3. The minimum mean standard deviation of an ensemble of five of the supervised classifiers is approximately 0.249, which is significantly more than the mean standard deviation of the $L^2$ score for the PU classifiers, 0.0598. This supports an ensemble of PU classifiers being more suitable for the method described in Section 3.3.6 to identify repeater candidates. Additionally, we note that the mean of the recall values for the PU classifiers is 0.8776, which exceeds the mean recall of any combination of the supervised classifiers, further supporting the increased ability of the PU ensemble to identify repeater candidates.

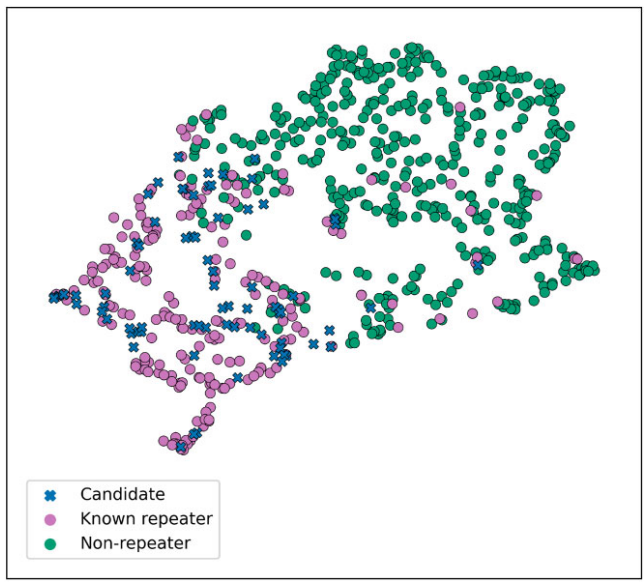

**Figure 1.** UMAP visualization of all sub-bursts from the merged data set, with candidates identified with PU classifiers highlighted. PU candidates appear to generally be clustered near known repeaters, which indicates a high degree of similarity in their features.

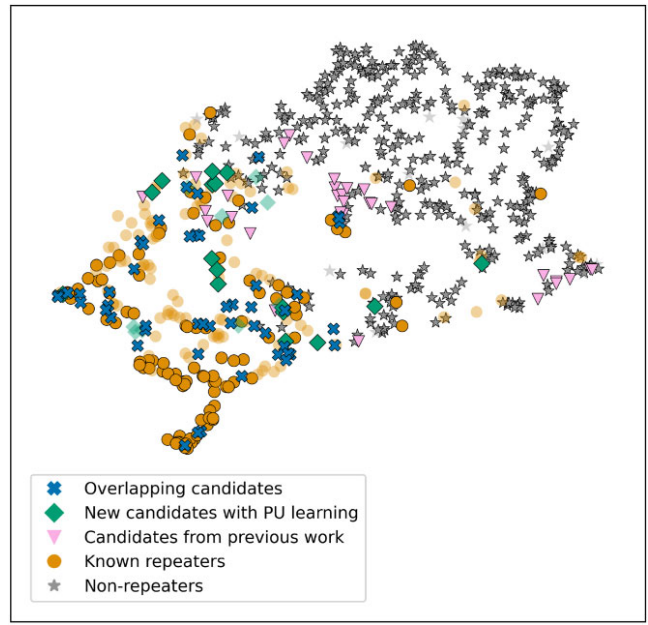

**Figure 2.** UMAP visualization of candidates identified by PU classifiers and candidates from previous work, with 2023 catalogue sub-bursts represented by borderless semitransparent points. There is a significant number of overlapping candidates, indicating high confidence in them being hidden repeaters. Compared to non-overlapping candidates from previous work, the PU classifier candidates are generally closer to either known repeaters or overlapping candidates.

## 4.2 Repeater candidates

66 apparent non-repeater bursts, consisting of 76 sub-bursts, are identified as candidate repeaters per the procedure in Section 3.3. A list of all the candidates is provided in Appendix C. The median number of times each candidate was flagged by a majority of the classifiers is 950, indicating a high degree of overall confidence in the results.

We use the Uniform Manifold Approximation and Projection (UMAP; McInnes et al. 2018a) technique to compare the candidates with other sub-bursts in a two-dimensional visualization. UMAP creates fuzzy connections between nearby points in high-dimensional space to produce a weighted graph and then uses a force-directed graph layout algorithm to project the data into lower dimensions. Since it constructs its graph based on local distances between points and therefore preserves local structures in the data, we may infer that a sub-burst is likely to be physically similar to its nearest neighbours in an UMAP plot. The global structure of the data should indicate differences in the populations of repeaters and non-repeaters. UMAP does not preserve global structure to the same extent as local structures, however, it has been found to be preferable in this aspect (McInnes et al. 2018a) over similar methods which prioritize local structure, such as t-SNE.

We implement an UMAP projection of the 13 input features from Section 2.2 using the `umap-learn` library (McInnes et al. 2018b). The 20 nearest neighbours of each sub-burst are used to create the graph. All other hyperparameters are left at their default values.

In Fig. 1, we plot the identified candidates alongside known repeaters and apparent non-repeaters in UMAP space. In Fig. 2, we compare our candidates with those identified by analysis of the 2021 catalogue by Pleunis et al. (2021), Luo et al. (2023), and Zhu-Ge et al. (2023). Of the 69 sub-bursts from the 2021 catalogue identified as candidates by the PU classifiers, 51 were also identified by at least one of the listed works. The other 18 originate from 18 distinct bursts, and are therefore considered to be 18 new repeating sources. 32 sub-bursts, from 28 sources are identified as candidates by any one of the three papers but are not identified by the PU classifiers.

Only 10 of the 32 sub-bursts from the silver sample (described in Section 2.1) are flagged as likely repeaters in our experiments. In Fig. 3, we compare the candidates identified in our experiments with the bursts from the silver sample described in Section 2.1. If we assume that the PU classifiers flagging a sub-burst of a burst as repeating, then the number of sub-bursts present in both the silver sample and the results of the PU ensemble increases by eight. We present the significance of this change in Fig. 4.

Sub-bursts from the silver sample that are not identified as candidate repeaters by the PU classifiers appear to be clustered closer to the apparent non-repeaters in Figs 3 and 4. We use the *k*-sample Anderson–Darling (AD) test with *k* = 20 to confirm the apparent differences in the features of each of the two sets of candidates compared to the sample known repeaters. We compare the features of the repeater candidate sub-bursts with the sample of known repeaters; and then compare sub-bursts from the silver sample with known repeaters. The results of the tests are presented in Table 8, where a *p*-value of 0.01 returned by the test implies > 99 per cent confidence that the two samples are drawn from different underlying distributions.

*p*-values are consistently lower for the features of bursts from the silver sample than for candidates identified by the PU classifiers. In particular, the difference in *p*-values is greater than two orders of magnitude for boxcar width, spectral index, rest-frame width, and frequency width, which are the four most important features distinguishing repeaters and non-repeaters as per the analysis in Section 4.3. This may be the reason for the relative lack of correspondence between candidates identified by our classifiers and those identified by Andersen et al. (2023).

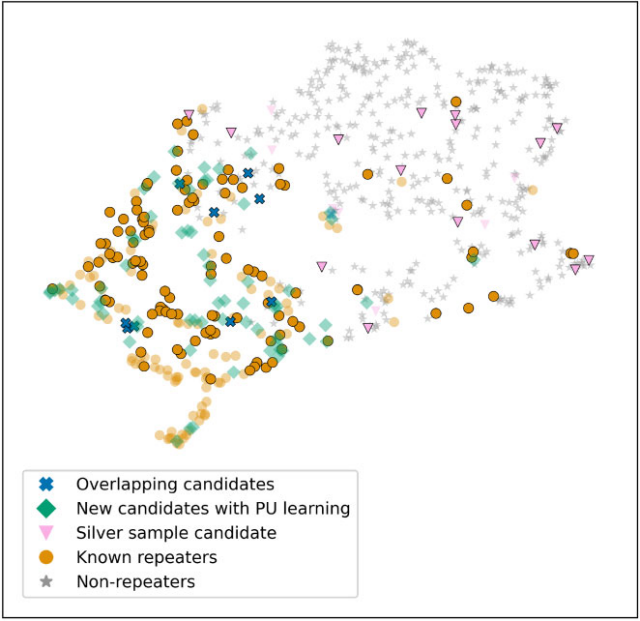

**Figure 3.** UMAP visualization of PU classifier candidates and the 2023 silver sample, with 2021 catalogue sub-bursts represented by borderless semitransparent points. There does not appear to be a high degree of correspondence between the two sets of candidates.

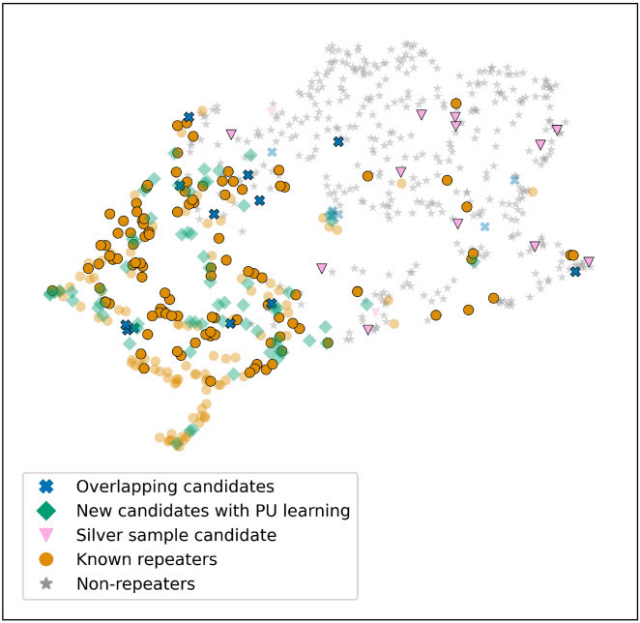

**Figure 4.** UMAP visualization of PU classifier candidates and the 2023 silver sample, now with all sub-bursts from any cluster with a PU classifier candidate all flagged as candidates. 2021 catalogue sub-bursts are represented by borderless semitransparent points.

**Table 8.** Anderson–Darling *p*-values denoting the probability of PU candidates and silver sample (SS) candidates originating from the same population as known repeaters. *p*-values are reported to three significant figures.

| Feature | $p_{PU}$ | $p_{SS}$ (all) | $p_{SS}$ (non-PU) |
|---|---|---|---|
| Right ascension | 0.004 | 0.006 | 0.03 |
| Signal-to-noise ratio | 0.01 | 0.04 | 0.04 |
| Extragalactic DM (NE2001) | 0.17 | 0.13 | 0.02 |
| Boxcar width | 0.24 | <.001 | <.001 |
| Flux | 0.13 | 0.02 | 0.0025 |
| Fluence | 0.13 | 0.10 | 0.21 |
| Spectral index | 0.01 | <.001 | <.001 |
| Spectral running | 0.02 | <.001 | <.001 |
| Rest-frame width | 0.08 | <.001 | <.001 |
| Peak frequency | 0.25 | 0.07 | 0.05 |
| Frequency width | 0.25 | <.001 | <.001 |
| Brightness temperature | 0.12 | <.001 | <.001 |
| Burst energy | 0.25 | 0.25 | 0.20 |

### 4.3 Feature importance

Evaluating the correlation between feature values and the predictions of the classifier ensemble can provide insights into the physical features which appear to distinguish between repeaters and non-repeaters. By shifting bursts which are likely to be hidden repeaters to the correct class, this study is likely to be more robust than a direct comparison with known labels.

For this, a mean measure of feature importance across the predictions of all five classifiers must be used. However, none of the five PU classifiers share a standardized metric for feature importance, because of which a model-agnostic method is required. Therefore, we employ the SHAP (SHapley Additive exPlanations) technique developed by Lundberg & Lee (2017), an explainable AI framework that utilizes the concept of Shapley values from game theory (Shapley et al. 1953). Shapley values measure the average expected marginal impact of each of the players of a game on the output of a game. In the context of evaluating classifiers, SHAP values measure the marginal contribution of each feature value to a corresponding set of predictions.

We use SHAP analysis to explain the mean value of the predictions for each of the five PU classifiers for all sub-bursts in data. Because of the computationally intense nature of SHAP analysis, we only use the PU classifiers from the 1000th train-test split.

#### 4.3.1 Absolute feature importance

In Fig. 5, we plot the mean absolute SHAP value for each feature for the 1000th iteration of each of the five PU classifiers. The mean of the five values is additionally noted to the right of the bars. Spectral index and frequency width are identified as the most important features by a significant margin. Rest-frame width and boxcar width also appear to be significant features.

#### 4.3.2 Trends in features corresponding to repeaters

While the absolute value of SHAP values is depicted in Fig. 5, their true values may be positive or negative. A positive SHAP value of a feature for a particular prediction signifies that its contribution was towards the positive class, i.e. that particular feature value increased the probability of the output being positive. The sum of the SHAP values for all of the feature values of an example will always add up to the final prediction of the classifiers for that example (Lundberg &

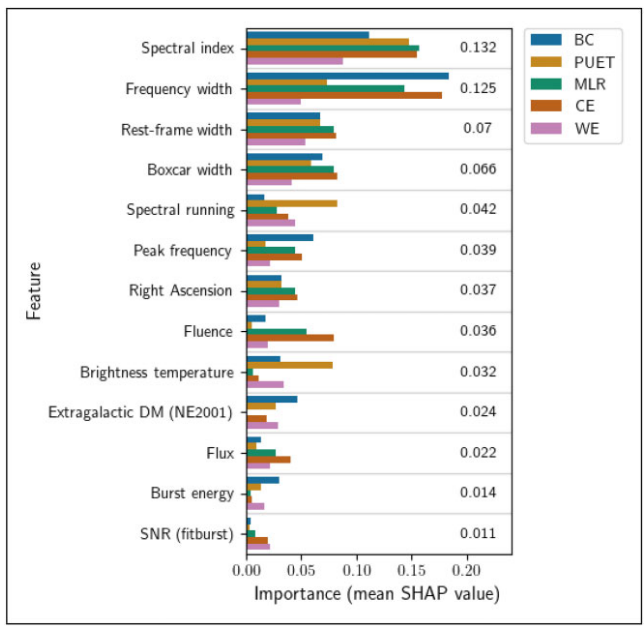

**Figure 5.** SHAP absolute feature importances for each PU classifier and mean importances across all classifiers. Trends in importance values appear are generally consistent across classifiers.

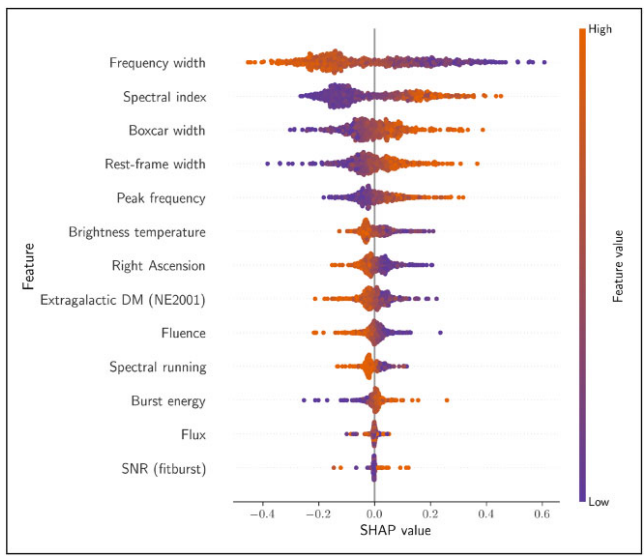

**Figure 6.** SHAP values for the mean predictions of the entire classifier ensemble. Fig. 5.

Lee 2017). Across an entire data set, this allows us to measure not only the magnitude of the importance of a feature in predicting the repeating nature of a burst but also the direction in which that feature must change to increase the probability of a burst being a repeater.

We present this analysis in Fig. 6. On each feature row, each dot represents a sub-burst. The $x$-position of the dot corresponds to the SHAP value of that feature for that sample. Dots are added along each feature row to show the density of predictions with that SHAP value for the feature. The colour represents the relative value of the feature itself.

From Figs 5 and 6, we can infer the following regarding the trends in feature values which appear to be characteristic of bursts from repeating sources.

(i) Spectral index is consistently an extremely significant feature, with repeat bursts appearing to have higher spectral indices. This relationship appears to be less prominent in previous results, with Zhong et al. (2022) only identifying marginal differences in the distributions; and Pleunis et al. (2021) identifying spectral running, a correlated feature, as statistically differing between repeaters and apparent non-repeaters.

(ii) Lower frequency width values correlate with a higher probability of a burst being a repeater, which corresponds with trends identified by Pleunis et al. (2021).

(iii) Repeaters appear to have longer temporal widths, both in terms of boxcar width (for the entire burst) and rest-frame width (for individual sub-bursts). This corresponds with previous work such as Connor, Miller & Gardenier (2020).

(iv) Peak frequency appears to have a non-negligible impact on the predictions, with higher peak frequency being indicative of the repeating nature of a burst. Zhu-Ge et al. (2023) and Chen et al. (2022) also found this feature to be significant.

(v) Brightness temperature is not identified as a particularly important feature on average; however, repeat bursts generally appear to have lower brightness temperature, with some exceptions. Considering each classifier individually from Fig. 5, it only appears to have a significant impact on the output of the PUET classifier. In contrast, Luo et al. (2023) and Zhu-Ge et al. (2023) found brightness temperature to be one of the most salient features differentiating repeaters and non-repeaters.

(vi) Repeaters appear to have higher flux values; however, flux measurements appear to have negligible impacts on the predictions in general. Similarly, extragalactic DM seems to have a negligible impact on the predictions, but repeaters appear to have lower extragalactic DM values.

(vii) Spectral running and burst energy appear to have negligible impacts and display no clearly contrasting trends between repeaters and non-repeaters.

## 5 CONCLUSIONS

Repeating sources of FRBs help offer valuable insights about the nature and origin of this enigmatic astronomical phenomenon. Identifying repeater candidates, i.e. apparently one-off FRB sources that are likely to be repeating in nature increases the efficiency of observation campaigns that survey known burst sky locations for follow-up events. Machine learning classifiers may be used to identify repeater candidates but the PU nature of FRB data, wherein only known repeaters are available and it is unknown which apparent non-repeaters are truly repeating and truly non-repeating, necessitate modifications to conventional supervised classification techniques.

In this work, we identified the field of positive unlabelled machine learning as being applicable to the identification of FRB repeater candidates. We used a data set of 643 FRBs observed by CHIME, consisting of 750 sub-bursts, to train eight conventional supervised classification techniques and optimized them to maximize a PU-specific performance metric. These classifiers were then used as a basis for the training of five PU-specific classification techniques, which were then optimized by the same method. Predictions from the five classifiers were extracted and combined over 1000 training iterations to produce a list of candidates that were flagged by a majority of the five classifiers in more than 10 per cent of iterations.

These experiments resulted in the identification of 66 candidate repeaters (Appendix C), of which 18 from the 2021 catalogue were available to but unidentified by previous works. UMAP comparisons further supported the overlapping candidates and newly identified candidates being physically similar to known repeaters. Using SHAP analysis, we found that the features identified by the ensemble as important characteristics of repeaters, such as lower frequency width and longer temporal width, were generally consistent with previous research. However, the PU ensemble most notably diverged from previous techniques in identifying high spectral indices as being the most important feature of repeaters and finding low brightness temperature to be a less significant feature.

The view that repeaters and non-repeaters have innate physical differences is further supported by the high degree of confidence with which PU learning techniques identified repeater candidates. These results may further support the existence of distinct physical source mechanisms for the origin of repeaters and non-repeaters; therefore, cataclysmic models for one-off FRBs cannot be completely eliminated from consideration, and progenitor models which include different mechanisms explaining repeater and one-off source behaviour may be required.

Apparent non-repeaters flagged as candidates by multiple previous works as well as by the PU ensemble with a high degree of confidence are the most likely sources to be hidden repeaters. Shifting some of these candidates to the sample of repeaters may reduce the contamination effect of hidden repeaters in future statistical comparisons of the two burst populations. Additionally, we recommend these candidates as ideal targets for future follow-up observations to maximize the chance of discovering new repeating FRBs.

## ACKNOWLEDGEMENTS

The authors are grateful for the valuable feedback provided by the anonymous referee regarding the manuscript. We would additionally like to thank the Cambridge Centre for International Research for facilitating this research work.

## DATA AVAILABILITY

All the data used in our experiments are sourced from the CHIME/FRB catalogue (chime-frb.ca). We utilize an adapted version of the code provided by Zhu-Ge et al. (2023) to extract secondary features of FRBs (github.com/JiamingZhang/FRB_ML.unsp). The code used by our experiments and the data generated are available at github.com/ArjunS07/pu-learning-for-frbs-2023.

## SUPPORTING INFORMATION

Supplementary data are available at *MNRAS* online.

**suppl_data**

Please note: Oxford University Press is not responsible for the content or functionality of any supporting materials supplied by the authors. Any queries (other than missing material) should be directed to the corresponding author for the article.

## APPENDIX A: ALL PARAMETERS USED FOR SUPERVISED CLASSIFIERS

| Model | Parameter | Values tested | Value used |
|---|---|---|---|
| **SVM** | C | 0.010 to 100 | 8.472 |
| | degree | 1 to 8 | 5 |
| | kernel | rbf | rbf |
| | gamma | gamma | gamma |
| | probability | False | False |
| | tol | $10^{-3}$ | $10^{-3}$ |
| | cache_size | 200 | 200 |
| | max_iter | $-1$ | $-1$ |
| | decision_function_shape | ovr | ovr |
| | break_ties | False | False |
| **LDA** | solver | svd, lsqr, eigen | lsqr |
| | store_covariance | True, False | True |
| | tol | $10^{-5}$ to $10^{-3}$ | $2.05 \times 10^{-5}$ |
| | shrinkage, priors, n_components, covariance_estimator | None | None |
| **LR** | tol | $10^{-5}$ to $10^{-3}$ | $5.62 \times 10^{-5}$ |
| | C | 0.010 to 100 | 63.500 |
| | solver | liblinear, newton-cholesky | liblinear |
| | max_iter | 100 to 500 | 162 |
| | penalty | L2 | L2 |
| | dual | False | False |
| | fit_intercept | True | True |
| | intercept_scaling | 1 | 1 |
| | class_weight | None | None |
| | multi_class | auto | auto |
| **Random forest** | n_estimators | 50 to 500 | 245 |
| | min_samples_split | 2 to 32 | 21 |
| | min_samples_leaf | 1 to 32 | 10 |
| | criterion | gini, entropy | entropy |
| | min_weight_fraction_leaf | 0 | 0 |
| | max_features | sqrt | sqrt |
| | max_leaf_nodes | None | None |
| | min_impurity_decrease | 0 | 0 |
| | bootstrap | True | True |
| | warm_start | False | False |
| | oob_score | False | False |
| | class_weight | None | None |
| | ccp_alpha | 0 | 0 |
| | max_samples | None | None |
| **AdaBoost** | n_estimators | 50 to 500 | 379 |
| | learning_rate | $10^{-3}$ to 1 | 0.280 |
| | algorithm | SAMME, SAMME.R | SAMME.R |
| | estimator | None | None |

*continued*

| Model | Parameter | Values tested | Value used |
|---|---|---|---|
| **XGBoost** | n_estimators | 50 to 500 | 472 |
| | eta | $10^{-3}$ to 10 | 4.46 |
| | gamma | $10^{-3}$ to 10 | $1.58 \times 10^{-2}$ |
| | min_child_weight | $10^{-3}$ to 10 | $1.32 \times 10^{-3}$ |
| | max_delta_step | $10^{-3}$ to 10 | $2.15 \times 10^{-2}$ |
| | max_leaves | 2 to 256 | 165 |
| | max_bin | 2 to 256 | 81 |
| | subsample | 0.1 to 1 | 0.809 |
| | colsample_bytree | 0.1 to 1 | 0.426 |
| | booster | gbtree | gbtree |
| | max_depth | 6 | 6 |
| | sampling_method | uniform | uniform |
| | colsample_bylevel | 1 | 1 |
| | colsample_bynode | 1 | 1 |
| | lambda | 1 | 1 |
| | alpha | 1 | 1 |
| | tree_method | auto | auto |
| | scale_pos_weight | 1 | 1 |
| | refresh_leaf | 1 | 1 |
| | process_type | default | default |
| | grow_policy | depthwise | depthwise |
| **LGBM** | n_estimators | 50 to 500 | 314 |
| | learning_rate | $10^{-3}$ to 1 | 0.763 |
| | subsample | 0.1 to 1 | 0.552 |
| | colsample_bytree | 0.1 to 1 | 0.875 |
| | boosting_type | gbdt | gbdt |
| | num_leaves | 31 | 31 |
| | max_depth | −1 | −1 |
| | subsample_for_bin | 200000 | 200000 |
| | objective | None | None |
| | class_weight | None | None |
| | min_split_gain | 0 | 0 |
| | min_child_weight | $10^{-3}$ | $10^{-3}$ |
| | min_child_samples | 20 | 20 |
| | subsample_freq | 0 | 0 |
| | reg_alpha | 0 | 0 |
| | reg_lambda | 0 | 0 |
| **Decision Tree** | min_samples_split | 2 to 32 | 20 |
| | min_samples_leaf | 1 to 32 | 2 |
| | criterion | gini, entropy | gini |
| | splitter | best | best |
| | max_depth | None | None |
| | min_weight_fraction_leaf | 0 | 0 |
| | max_features | None | None |
| | max_features | None | None |
| | min_impurity_decrease | None | None |
| | class_weight | None | None |
| | ccp_alpha | 0 | 0 |

# APPENDIX B: ALL PARAMETERS USED FOR PU CLASSIFIERS

CE, WE, and BC are implemented using the PULEARN (pulearn.github.io/pulearn) library. MLR and PUET are implemented with custom forks of the code provided by their original authors.

| Model | Parameter | Values tested | Optimal value |
|-------|-----------|---------------|---------------|
| CE | `estimator` | LDA, SVM, LR | LDA |
| | `holdout_ratio` | 0.1 to 0.8 | 0.673 |
| WE | `estimator` | LDA, SVM, LR | LR |
| | `holdout_ratio` | 0.1 to 0.8 | 0.387 |
| BC | `estimator` | LDA, SVM, LR | SVM |
| | `n_estimators` | 25 to 200 | 156 |
| | `max_samples` | 0.100 to 1 | 0.610 |
| | `max_features` | 0.100 to 1 | 0.977 |
| | `bootstrap` | True | True |
| | `bootstrap_features` | False | False |
| | `oob_score` | True | True |
| | `warm_start` | False | False |
| MLR | `learning_rate` | $10^{-4}$ to $10^{-1}$ | $1.57 \times 10^{-2}$ |
| | Epochs | 100 | 100 |
| PUET | `n_estimators` | 25 to 200 | 95 |
| | `risk_estimator` | uPU, nnPU | uPU |
| | `loss` | quadratic, logistic | quadratic |
| | `min_samples_leaf` | 1 to 10 | 3 |
| | `max_features` | sqrt, all | sqrt |
| | `max_candidates` | 1 to 10 | 5 |
| | `max_depth` | None | None |

## APPENDIX C: LIST OF REPEATER CANDIDATES

Below, we list the apparent non-repeater sub-bursts observed by CHIME/FRB which were identified as likely to be repeaters by PU classifiers across 1000 train-test splits with the highest confidence. Since each sub-burst was treated as a different example in our experiments, we identify each candidate using its TNS name as well as the specific sub-burst that was flagged by the classifier ensemble. For each sub-burst, we note the number of times it was flagged as a candidate. If the sub-burst was from catalogue 2, we note whether it was identified as a candidate in the 'silver sample' by Andersen et al. (2023). If the sub-bursts was from catalogue 1, we note if it was also identified as a candidate by **1**: Pleunis et al. (2021), **2**: Luo et al. (2023), or **3**: Zhu-Ge et al. (2023) if it was from catalogue 1. The entire list of candidates is available in the supplemental materials.

| TNS Name | Sub-burst | Count | Silver sample | Overlapping |
|----------|-----------|-------|---------------|-------------|
| FRB20190527A | 1 | 1000 | ✗ | 1, 2, 3 |
| FRB20190527A | 0 | 1000 | ✗ | 1, 2, 3 |
| FRB20190422A | 1 | 1000 | ✗ | 1, 2, 3 |
| FRB20190422A | 0 | 1000 | ✗ | 1, 2, 3 |
| FRB20190617B | 0 | 1000 | ✗ | 2, 3 |
| FRB20190429B | 0 | 1000 | ✗ | 2, 3 |
| FRB20190423B | 1 | 1000 | ✗ | 2, 3 |
| FRB20190423B | 0 | 1000 | ✗ | 2, 3 |
| FRB20190410A | 0 | 1000 | ✗ | 2, 3 |
| FRB20190329A | 0 | 1000 | ✗ | 2, 3 |
| FRB20190218B | 0 | 1000 | ✗ | 2, 3 |

This paper has been typeset from a TEX/LATEX file prepared by the author.